\documentclass[a4paper,american,orivec,fleqn]{llncs}
\usepackage{amsfonts}
\usepackage{amsmath}
\usepackage{mathtools}[2011/02/12]
\usepackage{mdwlist}
\usepackage{stmaryrd}
\usepackage[fancyproofs,fancyexamples,squareitemtag]{theorems}[2015/07/09]

\providecommand*{\parensmathoper}[3][]{\ensuremath{\mathoper{#2}\ifempty{#3}{}{#1(#3#1)}}}
\providecommand*{\monobioperator}[3]{
  \newcommand*{#1}[2]{{\ifempty{##2}{#3##1}{##1#2##2}}}
}
\newcommand*{\C}{\mathbb{C}}

\newcommand*{\dfn}{\coloneq}
\newcommand*{\wrt}{with respect to}
\providecommand*{\mathoper}[1]{\mathop{\mathit{#1}}\nolimits}
\providecommand*{\true}{\mathit{true}}   
\providecommand*{\false}{\mathit{false}} 
\providecommand*{\ie}   {i.e.,} 
\providecommand*{\resp} {respectively}

\makeatletter
\providecommand{\ifempty}[3]{\def\@@@temp{#1}\ifx\@@@temp\@empty#2\else#3\fi}
\@ifpackageloaded{mathtools}{}{\PackageWarning{WGmacros}
{Please load `mathtools' package explicitely in preamble}
\RequirePackage{mathtools}}
\makeatother

\newcommand*{\Osyms}{\Omega}
\newcommand*{\FOsyms}{\widetilde\Osyms}    
\newcommand*{\Fsyms}{\Sigma}

\makeatletter
\@ifpackageloaded{MnSymbol}{}{\PackageWarning{WGmacros}%
{Please load `MnSymbol' package explicitely in preamble}
\RequirePackage{MnSymbol}}
\makeatother

\newcommand*{\N}{\ensuremath{\mathbb{N}}}
\newcommand*{\R}{\ensuremath{\mathbb{R}}}
\newcommand*{\Z}{\ensuremath{\mathbb{Z}}}
\newcommand*{\Fp}{\ensuremath{\mathbb{F}}}
\newcommand*{\real}{\parensmathoper{\mathit{\chi_{r}}}}

\newcommand*{\ra}{\rightarrow}
\newcommand*{\FPtoR}{\parensmathoper{{R}}}
\newcommand*{\RtoF}[2][]{\mathit{F}_{#1}\ifempty{2}{}{(#2)}}
\newcommand*{\FtoR}{\parensmathoper{\mathit{R}}}
\newcommand*{\FtoRB}{\parensmathoper{\mathit{R}_{\FBExprDom}}}
\newcommand*{\FtoRA}{\parensmathoper{\mathit{R}_{\FAExprDom}}}
\newcommand*{\AExpr} {\mathit{A}}
\newcommand*{\BExpr} {\mathit{B}}
\newcommand*{\FAExpr}{\widetilde{\mathit{A}}}
\newcommand*{\FBExpr}{\widetilde{\mathit{B}}}
\newcommand*{\AExprDom} {\mathbb{A}}
\newcommand*{\BExprDom} {\mathbb{B}}
\newcommand*{\FAExprDom}{\widetilde{\mathbb{A}}}
\newcommand*{\FBExprDom}{\widetilde{\mathbb{B}}}

\newcommand*{\nleqB}{\mathrel{\not\Rightarrow}}
\newcommand*{\FStmDom}{\mathbb{S}} 
\newcommand*{\FStm}{\mathit{S}}
\newcommand*{\Prog}{\mathbb{P}}
\newcommand*{\supC}{\mathbf{C}}
\monobioperator{\lubC}{\sqcup}{\bigsqcup}
\newcommand*{\topC}{\supC}
\newcommand*{\fpcnst}{\mathit{\tilde{d}}}
\newcommand*{\rcnst}{\mathit{d}}
\newcommand*{\FExpr}{\FStm}
\newcommand{\bexpr}{\phi}
\newcommand{\vraexpr}{\mathit{expr}}
\newcommand{\vfpaexpr}{\widetilde{\vraexpr}}
\newcommand*{\fprog}{P}
\newcommand*{\fpfun}{\tilde{f}}

\newcommand*{\fpaexpr}{\tilde{A}}
\newcommand*{\stm}{S}
\newcommand*{\rop}{\ensuremath{\mathit{op}}}
\newcommand*{\fpop}{\ensuremath{\widetilde{\rop}}}
\newcommand*{\taglet}{\mathrel{\mathit{let}}}
\newcommand*{\tagin}{\mathrel{\mathit{in}}}
\newcommand*{\tagif}{\mathrel{\mathit{if}}}
\newcommand*{\tagthen}{\mathrel{\mathit{then}}}
\newcommand*{\tagelse}{\mathrel{\mathit{else}}}
\newcommand*{\tagelsif}{\mathrel{\mathit{elsif}}}

\newcommand*{\letStm}[3]{\taglet #1=#2 \tagin #3}
\newcommand*{\ite}[3]{\tagif #1 \tagthen #2 \ifempty{#3}{}{\tagelse #3}}

\newcommand*{\stabWarning}{\mathit{warning}}
\newcommand*{\Var}{\mathbb{V}}
\newcommand*{\FVar}{\widetilde{\mathbb{V}}}
\newcommand*{\Env}{\mathit{Env}}
\newcommand*{\env}{\nu}
\newcommand{\fvar}[1][x]{\tilde{#1}}
\newcommand*{\botEnv}{\bot_{\Env}}
\newcommand*{\botUnstt}{\bot_{\unstt}}
\newcommand*{\propGuard}[3]{\ifempty{#1}{\Downarrow}{\ifempty{#2}{\Downarrow_{\cond{#1}{#2}}} {{#3}\Downarrow_{\cond{#1}{#2}}}}}

\newcommand*{\cond}[2]{(#1, #2)}
\newcommand*{\ct}[5]{\langle #1, #2 \rangle_{#5} \twoheadrightarrow (#3, #4)}
\newcommand*{\rcond}{\eta}
\newcommand*{\fpcond}{\tilde{\eta}}
\newcommand*{\rres}{r}
\newcommand*{\fpres}{\tilde{r}}
\newcommand*{\stt}{\mathbf{s}}
\newcommand*{\unstt}{\mathbf{u}}
\newcommand*{\Ssem}[3]{\mathoper{\mathcal{E}}\ifempty{#1}{}{\llbracket#1\rrbracket_{#2}}}
\newcommand*{\emin}{e_{\mathit{min}}}
\newcommand*{\rv}[1]{\mathit{r}\ifempty{#1}{}{_{#1}}}

\newcommand*{\fpv}[1]{\tilde{\mathit{v}}\ifempty{#1}{}{_{#1}}}
\newcommand*{\Rfpv}[1]{\FPtoR{\tilde{\mathit{v}}\ifempty{#1}{}{_{#1}}}}
\newcommand*{\fpexp}[1]{\ensuremath{\mathit{e}_{#1}}}

\newcommand*{\fpbool}{\tilde{\phi}}
\newcommand*{\precisa}{PRECiSA}

\newcommand*{\PVS}{PVS}
\newcommand*{\FPTuner}{FPTuner}

\newcommand{\evalBExpr}[2]{\mathit{eval}_{\BExprDom}(#1,#2)}
\newcommand{\evalFBExpr}[2]{\widetilde{\mathit{eval}}_{\FBExprDom}(#1,#2)}
\newcommand*{\betaPos}{\parensmathoper{\beta^{+}}}
\newcommand*{\betaNeg}{\parensmathoper{\beta^{-}}}
\newcommand*{\tauProg}{\parensmathoper{\tau}}
\newcommand*{\rerr}{\epsilon}
\newcommand*{\fperr}{\rerr}
\newcommand*{\fv}{\parensmathoper{\mathit{fv}}}

\begin{document}

\title{Eliminating Unstable Tests in Floating-Point
Programs}

\author{Laura Titolo\inst{1}  \and C\'{e}sar A. Mu\~{n}oz\inst{2} \and Marco A. Feli\'{u}\inst{1} \and Mariano M. Moscato\inst{1}}

\institute{National Institute of Aerospace,\\
\email{\{laura.titolo,marco.feliu,mariano.moscato\}@nianet.org}
\thanks{Research by the first, the third, and the fourth authors was supported by the
National Aeronautics and Space Administration under NASA/NIA Cooperative Agreement NNL09AA00A.}
\and
NASA Langley Research Center,\\
\email{cesar.a.munoz@nasa.gov}
}

\maketitle

\begin{abstract}
Round-off errors arising from the difference between real numbers and their floating-point representation cause the control flow of conditional floating-point statements to deviate from the ideal flow of the real-number computation.
This problem, which is called test instability, may result in a significant difference between the  computation of a floating-point program and the expected output in real arithmetic.
In this paper, a formally proven program transformation is proposed to detect and correct the effects of unstable tests.
The output of this transformation is a floating-point program that is guaranteed to return either the result of the original floating-point program when it can be assured that both its real and its floating-point flows agree or a warning when these flows may diverge.
The proposed approach is illustrated with the transformation of the core computation of a polygon containment algorithm developed at NASA that is used in a geofencing system for unmanned aircraft systems.
\end{abstract}


\section{Introduction}
\label{sec:intro}

Floating-point numbers are widely used to represent real numbers in computer programs since they offer a good trade-off between efficiency and precision.
The round-off error of a floating-point expression is the difference between the ideal computation in real arithmetic and the actual floating-point computation.
These round-off errors accumulate during numerical computations.
Besides having a direct effect on the result of mathematical operations, round-off errors may significantly impact the control flow of a program.
This happens when the guard of a conditional statement contains a floating-point expression whose round-off error makes the actual Boolean value of the guard differ from the value that would be obtained assuming real arithmetic.
In this case, the conditional statement is called an {\em unstable test.}
Unstable tests are an inherent feature of floating-point programs.
In general, it is not possible to completely avoid them.
However, it is possible to mitigate their effect by transforming the original program into another program that conservatively (and soundly) detects and corrects unstable tests. 

This paper presents a program transformation technique to transform a given program into a new one that returns either the same result of the original program or a warning when the real and floating-point flows may diverge.
This transformation is parametric \wrt{} two Boolean abstractions that take into consideration the round-off error in the expressions occurring in the guard.
The transformation replaces the unstable conditions with more restrictive conditions that are guaranteed to preserve the control flow of stable tests.
The correctness of the proposed transformation is formally verified in the Prototype Verification System (PVS)~\cite{OwreRS92}.

The remainder of the paper is organized as follows.
\smartref{sec:fp_err} provides technical background on floating-point numbers and round-off errors.
The proposed program transformation technique is presented in \smartref{sec:transformation}. \smartref{sec:polycarp} illustrates this technique by transforming the core logic of an algorithm for polygon containment that is part of a geofencing system developed by NASA. 
\smartref{sec:related} discusses related work and \smartref{sec:concl} concludes the paper.

\section{Round-Off Errors and Unstable Tests}
\label{sec:fp_err}

A floating-point number can be formalized as a pair of integers $(m,\fpexp{}) \in \Z^2$, where $m$ is called the \emph{significand} and $\fpexp{}$ the \emph{exponent} of the float~\cite{BoldoMunoz06,Daumas2001}.
A floating-point \emph{format} $f$ is defined as a pair of integers $(p,\emin)$, where $p$ is called the {\em precision} and $\emin$ is called the \emph{minimal exponent}.
For instance, IEEE single and double precision floating-point numbers are specified by the formats $(24, 149)$ and $(53, 1074)$, \resp.
A \emph{canonical} float is a float such that is either a normal or subnormal.
A \emph{normal} float is a float such that the significand cannot be multiplied by the radix and still fit in the format.
A \emph{subnormal} float is a float having the minimal exponent such that its significand can be multiplied by the radix and still fit in the format.
Henceforth, $\Fp$ will denote the set of floating-point numbers in canonical form and the expression $\fpv{}$ will denote a floating-point number $(m,e)$ in $\Fp$.
A conversion function $\FPtoR{}: \Fp\rightarrow \R$ is defined to refer to the real number represented by a given float, \ie{} $\FPtoR{(m,\fpexp{})} = m \cdot \beta^{\fpexp{}}$.

The expression $\RtoF[f]{\rv{}}$ denotes the floating-point number in format $f$ \emph{closest} to $\rv{}$. The format $f$ will be omitted when clear from the context.
Let $\fpv{}$ be a floating-point number that represents a real number $\rv{}$, the difference $|\Rfpv{} - \rv{}|$ is called the \emph{round-off error} (or \emph{rounding error}) of $\fpv{}$ \wrt{} $\rv{}$.

\subsection{Unstable tests}

Given a set $\FOsyms$ of pre-defined floating-point operations, the corresponding set $\Osyms$ of operations over real numbers, a finite set $\Var$ of variables representing real values, and a finite set $\FVar$ of variables representing floating-point values, where $\Var$ and $\FVar$ are disjoint, the sets $\AExprDom$ and $\FAExprDom$ of arithmetic expressions over real numbers and over floating-point numbers, \resp, are defined by the following grammar.
\begin{align*}
&\AExpr  ::= \rcnst  \mid x \mid \rop(\AExpr,\ldots,\AExpr),\qquad\qquad \FAExpr ::= \fpcnst \mid \fvar \mid \fpop(\FAExpr,\ldots,\FAExpr),
\end{align*}
where $\AExpr \in \AExprDom$, $\rcnst \in \R$, $x \in \Var$, $\rop \in \Osyms$, $\FAExpr \in \FAExprDom$, $\fpcnst \in \Fp$, $\fvar \in \FVar$, $\fpop \in \FOsyms$.
It is assumed that there is a function $\real{}: \FVar \ra \Var$ that associates to each floating-point variable $\fvar$ a variable $x \in \Var$ representing the real value of $\fvar$.
The function $\FtoRA{} : \FAExprDom \rightarrow \AExprDom$ converts an arithmetic expression on floating-point numbers to an arithmetic expression on real numbers.
It is defined by simply replacing each floating-point operation with the corresponding one on real numbers and by applying $\mathit{R}$ and $\real{}$ to floating-point values and variables, respectively.

Boolean expressions are defined by the following grammar.
\begin{align*}
   \BExpr &::= \true \mid \false \mid \BExpr \wedge \BExpr
               \mid \BExpr \vee \BExpr \mid \neg \BExpr
               \mid \AExpr < \AExpr \mid \AExpr = \AExpr
               \mid \FAExpr < \FAExpr \mid \FAExpr = \FAExpr,
\end{align*}
where $\AExpr\in\AExprDom$ and $\FAExpr\in\FAExprDom$.
The conjunction~$\wedge$, disjunction~$\vee$, negation~$\neg$,
$\true$, and $\false$ have the usual classical logic meaning.
The symbols $\BExprDom$ and $\FBExprDom$ denote the domain of Boolean expressions over real
and floating-point numbers, \resp. 
The function $\FtoRB{} : \FBExprDom \rightarrow \BExprDom$ converts a Boolean expression on floating-point numbers to a Boolean expression on real numbers.
Given a variable assignment $\sigma: \Var \rightarrow \R$, $\evalBExpr{\sigma}{B} \in \{\true,\false\}$ denotes the evaluation of the real Boolean expression $B$.
Similarly, given $\widetilde{B}\in\FBExprDom$ and $\widetilde{\sigma}: \FVar \rightarrow \Fp$, $\evalFBExpr{\widetilde{\sigma}}{\widetilde{B}} \in \{\true,\false\}$ denotes the evaluation of the floating-point Boolean expression $\widetilde{B}$.

The expression language considered in this paper contains binary and
$n$-ary conditionals, let expressions, arithmetic expressions,
and a warning exceptional statement.
Given a set $\Fsyms$ of function symbols, the syntax of program expressions in $\FStmDom$ is given by the following grammar.
\begin{equation}
    \label{eq:lang}
\begin{aligned}
   \FStm ::= & \FAExpr \ \mid\ \ite{\FBExpr}{\FStm}{\FStm}
              \ \mid\ \tagif \FBExpr \tagthen \FStm\
              [\tagelsif \FBExpr \tagthen \FStm]_{i=1}^{n} \tagelse \FStm\\
		      & \mid\ \letStm{\fvar}{\FAExpr}{\FStm}
              \ \mid\ \stabWarning,
\end{aligned}
\end{equation}
where $\FAExpr\in\FAExprDom$, $\FBExpr\in\FBExprDom$, $\fvar \in \FVar$, and $n\in\N^{>0}$.
The notation $[\tagelsif \FBExpr \tagthen \FStm]_{i=1}^{n}$ denotes a list of $n$ $\tagelsif$ branches.

A program is a \emph{function declaration} of the form $\fpfun(\fvar_1,\ldots,\fvar_m)=\FExpr$, where $\fvar_1,\ldots,\fvar_m$ are pairwise distinct variables in $\FVar$ and all free variables appearing in $\FExpr$ are in $\{\fvar_1, \ldots,\fvar_m\}$.
The natural number $m$ is called the \emph{arity} of $\fpfun$.
The set of programs is denoted as $\Prog$.

When if-then-else guards contain floating-point expressions, the
output of the considered program is not only directly influenced by
rounding errors, but also by the error of taking the incorrect branch
in the case of unstable tests.
\begin{definition}[Conditional Instability]\label{def:stable}
    A function declaration $\fpfun(\!\fvar_1,\!\dots,\!\fvar_n\!)\! =\! \FStm$
    is said to have an \emph{unstable conditional} when its body 
    contains a conditional statement
    of the form $\ite{\tilde{\phi}}{\FStm_1}{\FStm_2}$ and there exist
    two assignments $\tilde\sigma: \{\fvar_1,\dots,\fvar_n\} \rightarrow \Fp$
    and $\sigma: \{\real{\fvar_1},\dots,\real{\fvar_n}\} \rightarrow \R$ such that for all
    $i\in\{1,\dots,n\}$,
    $\sigma(\real{\fvar_i}) = \mathit{R}({\tilde{\sigma}(\fvar_i)})$ and
    $\evalBExpr{{\sigma}}{\FtoRB{\tilde{\phi}}} \neq \evalFBExpr{\tilde\sigma}{\tilde{\phi}}$.
    Otherwise, the conditional expression is said to be \emph{stable}.
\end{definition}
In other words, a conditional statement (or test) $\tilde{\phi}$ is unstable when there exists an assignments from the free variables $\tilde{x}_i$ in $\tilde{\phi}$ to $\Fp$ such that $\tilde{\phi}$ evaluates to a different Boolean value with respect to its real valued counterpart $\FtoRB{\tilde{\phi}}$.
In these cases, the program is said to follow an \emph{unstable path}, otherwise, when the flows coincide, it is said to follow a \emph{stable path}.

\subsection{Floating-Point Denotational Semantics}

This section presents a compositional denotational semantics for the expression language of Formula~\eqref{eq:lang} that models both real and floating-point path conditions and outputs.
This semantics is a modification of the one introduced in \cite{MoscatoTDM17} and \cite{TitoloFMM18}. 
The proposed semantics collects for each combination of real and floating-point program paths: the real and floating-point path conditions, two symbolic expressions representing the value of the output assuming the use of real and floating-point arithmetic, \resp, and a flag indicating if the element refers to either a stable or an unstable path.
This information is stored in a \emph{conditional tuple}.
\begin{definition}[Conditional Tuple]\label{def:ceb}
    A \emph{conditional tuple} is an expression of the form
    $\ct{\rcond}{\fpcond}{\rres}{\fpres}{t}$, where
    $\rcond \in \BExprDom$,
    $\fpcond \in \FBExprDom$,
    $\rres \in \AExprDom\cup\{\botUnstt\}$,
    $\fpres \in \FAExprDom\cup\{\botUnstt\}$,
    and $t\in\{\stt, \unstt\}$.
\end{definition}
Intuitively, $\ct{\rcond}{\fpcond}{\rres}{\fpres}{t}$ indicates that if the condition $\rcond \wedge \fpcond$ is satisfied, the output of the ideal real-valued implementation of the program is $\rres$ and the output of the floating-point execution is $\fpres$.
The sub-index $t$ is used to mark by construction whether a conditional tuple corresponds to an unstable path, when $t = \unstt$, or to a stable path, when $t = \stt$.
The element $\botUnstt$ represents the output of the warning construct.
Let $\topC$ be the set of all conditional error bounds, and $\C \dfn \wp (\topC)$ be the domain formed by sets of conditional error bounds.

An \emph{environment} is defined as a function mapping a variable to a set of conditional tuples, \ie{} $\Env : \FVar \ra \C$.
The empty environment is denoted as $\botEnv$ and maps every variable to the empty set $\emptyset$.

Given $\env \in \Env$, the semantics of program expressions is defined in \smartref{fig:sem} as a function $\Ssem{}{}{}{} : \FStmDom \times \Env \ra \C$ that returns the set of conditional tuples representing the possible real and floating-point computations and their corresponding path conditions. The operator $\lubC{{}}{{}}$ denotes the least upper bound of the domain of conditional error bounds.
\begin{figure}[t!]
{\small
\begin{align*}
    %
    %
	&\Ssem{\fpcnst}{\env}{\I} \dfn
    \{\ct{\true}{\true}{\FtoR{\fpcnst}}{\fpcnst}{\stt}\}
    \\[1ex]
    %
    %
    &\Ssem{\stabWarning}{\env}{\I} \dfn \{\ct{\true}{\true}{\botUnstt}{\botUnstt}{\stt}\}
    \\[1ex]
    %
    %
	&\Ssem{\fvar}{\env}{\I}  \dfn 
    \begin{cases}
     \{\ct{\true}{\true}{\real{\fvar}}{\fvar}{\stt}\} &\text{if $\env(\fvar)=\emptyset$}\\
     \env(\fvar)
     &\text{otherwise}
     \end{cases}
     \\[1ex]
    %
    %
	&\Ssem{\fpop(\FAExpr_{i})_{i=1}^{n}}{\env}{\I}  \dfn
     \lubC{}{}\{
     \begin{aligned}[t]
     &\ct{\bigwedge_{i=1}^{n}\bexpr_i}
         {\bigwedge_{i=1}^{n}\tilde\bexpr_i}{\rop(r_i)_{i=1}^{n}}
	     {\fpop(\tilde{r}_i)_{i=1}^{n}}{\stt}
     \mid \forall 1 \le i \le n \colon \\
     &\ct{\bexpr_i}{\tilde{\bexpr}_i}{r_i}{\tilde{r}_i}{\stt}\in \Ssem{\FAExpr_i}{\env}{\I},
     \bigwedge_{i=1}^{n}\bexpr_i \nleqB \false,
     \bigwedge_{i=1}^{n}\tilde\bexpr_i \nleqB \false\}
     \end{aligned}
     \\[1ex]
    %
    %
    &\Ssem{\letStm{\fvar}{\FAExpr}{\FStm}}{\env}{\I}
      \dfn \Ssem{\FStm}{\env[\fvar \mapsto \Ssem{\FAExpr}{\env}{\I}]}{\I}
      \\[1ex]
    %
    %
    &\Ssem{\ite{\FBExpr}{\FStm_1}{\FStm_2}}{\env}{\I} \dfn
    \propGuard{\FtoRB{\FBExpr}}{\FBExpr}{\Ssem{\FStm_1}{\env}{\I}} \lubC{{}}{{}}\
    \propGuard{\neg\FtoRB{\FBExpr}}{\neg \FBExpr}{\Ssem{\FStm_2}{\env}{\I}} \lubC{{}}{{}} \\
    &\quad \lubC{}{}\{
        \ct{\bexpr_2}{\tilde{\bexpr}_1}{r_2}{\tilde{r}_1}{\unstt} \mid
        \ct{\bexpr_1}{\tilde{\bexpr}_1}{r_1}{\tilde{r}_1}{\stt}
        \in \Ssem{\FStm_1}{\env}{\I},\\
        &\qquad\qquad \ct{\bexpr_2}{\tilde{\bexpr}_2}{r_2}{\tilde{r}_2}{\stt}
        \in \Ssem{\FStm_2}{\env}{\I} \}
        \propGuard{\neg\FtoRB{\FBExpr}}{\FBExpr}{} \lubC{{}}{{}}
    \\
    &\quad \lubC{}{}\{
        \ct{\bexpr_1}{\tilde{\bexpr}_2}{r_1}{\tilde{r}_2}{\unstt} \mid
        \ct{\bexpr_1}{\tilde{\bexpr}_1}{r_1}{\tilde{r}_1}{\stt}
        \in \Ssem{\FStm_1}{\env}{\I},\\
        &\qquad\qquad \ct{\bexpr_2}{\tilde{\bexpr}_2}{r_2}{\tilde{r}_2}{\stt}
        \in \Ssem{\FStm_2}{\env}{\I}\}
        \propGuard{\FtoRB{\FBExpr}}{\neg \FBExpr}{}
    \\[1ex]
    %
    %
    &\Ssem{\tagif \FBExpr_1 \tagthen \FStm_1\
    [\tagelsif \FBExpr_i \tagthen \FStm_i]_{i=2}^{n-1} \tagelse \FStm_{n}}{\env}{\I} \dfn\\
    &\quad \lubC{}{}_{i=1}^{n-1} \propGuard{\FBExpr_i \wedge \bigwedge_{j=1}^{i-1}\neg\FBExpr_j}
    {\FtoR{\FBExpr_i} \wedge \bigwedge_{j=1}^{i-1}\neg\FtoR{\FBExpr_j}}
    {\Ssem{\FStm_i}{\env}{\I}}\\
    &\quad\lubC{{}}{{}}
    \propGuard{\bigwedge_{j=1}^{n-1}\neg\FBExpr_j}
    {\bigwedge_{j=1}^{n-1}\neg\FtoR{\FBExpr_j}}
    {\Ssem{\FStm_n}{\env}{\I}} \lubC{{}}{{}}\\
    &\quad {}\lubC{}{} \{
        \ct{\eta_i}{\tilde{\eta}_j}{r_i}{\tilde{r}_j}{\unstt} \mid
        i,j \in \{1,\dots, n-1\}, i \neq j,\
        \ct{\eta_i}{\tilde{\eta}_i}{r_i}{\tilde{r}_i}{\stt}
        \in \Ssem{\FStm_i}{\env}{\I},\\
        &\qquad\qquad\ct{\eta_j}{\tilde{\eta}_j}{r_j}{\tilde{r}_j}{\stt}
        \in \Ssem{\FStm_j}{\env}{\I}\}
        \propGuard{\FBExpr_j \wedge \bigwedge_{k=1}^{j-1}\neg\FBExpr_k}
                  {\FtoR{\FBExpr_i} \wedge \bigwedge_{k=1}^{i-1}\neg\FtoR{\FBExpr_k}}{}    
    \lubC{{}}{{}}\\
    &\quad \lubC{}{} \{
        \ct{\eta_i}{\tilde{\eta}_n}{r_i}{\tilde{r}_n}{\unstt} \mid
        i \in \{1,\dots, n-1\},\
        \ct{\eta_i}{\tilde{\eta}_i}{r_i}{\tilde{r}_i}{\stt}
        \in \Ssem{\FStm_i}{\env}{\I},\\
        &\qquad\qquad\ct{\eta_n}{\tilde{\eta}_n}{r_n}{\tilde{r}_n}{\stt}
        \in \Ssem{\FStm_n}{\env}{\I}\}
        \propGuard{\bigwedge_{k=1}^{n-1}\neg\FBExpr_k}
                  {\FtoR{\FBExpr_i} \wedge \bigwedge_{k=1}^{i-1}\neg\FtoR{\FBExpr_k}}{}
    \lubC{{}}{{}}\\
    &\quad \lubC{}{} \{
        \ct{\eta_n}{\tilde{\eta}_i}{r_n}{\tilde{r}_i}{\unstt} \mid
        i \in \{1,\dots, n-1\},\
        \ct{\eta_i}{\tilde{\eta}_i}{r_i}{\tilde{r}_i}{\stt}
        \in \Ssem{\FStm_i}{\env}{\I},\\
        &\qquad\qquad \ct{\eta_n}{\tilde{\eta}_n}{r_n}{\tilde{r}_n}{\stt}
        \in \Ssem{\FStm_n}{\env}{\I}\}
        \propGuard{\FBExpr_i \wedge \bigwedge_{k=1}^{i-1}\neg\FBExpr_k}
                  {\bigwedge_{k=1}^{n-1}\neg\FtoR{\FBExpr_k}}{}
\end{align*}
}
\caption{Semantics of a program expression.}
\label{fig:sem}
\end{figure}

The semantics of a variable $\fvar\in\FVar$ consists of two cases. If $\fvar$ belongs to the environment, then the variable has been previously bound to a program expression $\FStm$ through a let-expression. 
In this case, the semantics of $\fvar$ is exactly the semantics of $\FStm$.
If $\fvar$ does not belong to the environment, then $\fvar$ is a parameter of the function.
Here, a new conditional error bound is added with a placeholder $\real{\fvar}$ representing the real value of $\fvar$.
The semantics of a floating-point arithmetic operation $\fpop$ is computed by composing the semantics of its operands.
The real and floating-point values are obtained by applying the corresponding arithmetic operation to the values of the operands.
The new conditions are obtained as the combination of the conditions of the operands.
The semantics of the expression $\letStm{\fvar}{\FAExpr}{\FStm}$ updates the current environment by associating with variable $\fvar$ the semantics of expression $\FAExpr$.

The semantics of the conditional $\ite{\FBExpr}{\FStm_1}{\FStm_2}$ uses an auxiliary operator $\propGuard{}{}{}$.
\begin{definition}[Condition propagation operator]
    Given $b\in \BExprDom$ and $\tilde{b} \in \FBExprDom$,
    $\propGuard{b}{\tilde{b}}{\ct{\bexpr}{\tilde{\bexpr}}{\rres}{\fpres}{t}} =
    \ct{\bexpr\wedge b}{\tilde{\bexpr}\wedge \tilde{b}}{\rres}{\fpres}{t}$
    if $\phi \wedge b \wedge \tilde{\phi}\wedge\tilde{b} \nleqB \false$,
    otherwise it is undefined. The definition of
    $\propGuard{}{}{}$ naturally extends to
    sets of conditional tuples: given $C\in\C$, 
    $\propGuard{b}{\tilde{b}}{C} = \bigcup_{c \in C} \propGuard{b}{\tilde{b}}{c}$.
\end{definition}
The semantics of $\FStm_1$ and $\FStm_2$ are enriched with the information about the fact that real and floating-point control flows match, \ie{} both $\FBExpr$ and $\FtoRB{\FBExpr}$ have the same value.
In addition, new conditional tuples are built to model the unstable cases when real and floating-point control flows do not coincide and, therefore, real and floating-point computations diverge.
For example, if $\FBExpr$ is satisfied but $\FtoRB{\FBExpr}$ is not, the $\mathit{then}$ branch is taken in the floating-point computation, but the $\mathit{else}$ would have been taken in the real one.
In this case, the real condition and its corresponding output are taken from the semantics of $\FStm_2$, while the floating-point condition and its corresponding output are taken from the semantics of $\FStm_1$.
The condition $\cond{\neg\FtoRB{\FBExpr}}{\FBExpr}$ is propagated in order to model that $\FBExpr$ holds but $\FtoRB{\FBExpr}$ does not.
The conditional tuples representing this case are marked with $\unstt$.

Similarly, the semantics of an n-ary conditional is composed of stable and unstable cases.
The stable cases are built from the semantics of all the program sub-expressions $S_i$ by enriching them with the information stating that the correspondent guard and its real counter-part hold and all the previous guards and their real counterparts do not hold.
All the unstable combinations are built by combining the real parts of the semantics of a program expression $S_i$ and the floating-point contributions of a different program expression $S_j$.  
In addition, the operator $\propGuard{}{}{}$ is used to propagate the information that the real guard of $S_i$ and the floating-point guard of $S_j$ hold, while the guards of the previous branches do not hold.

\section{Program Transformation}
\label{sec:transformation}

In this section, a program transformation is proposed for detecting when round-off errors affect the evaluation of floating-point conditionals and for ensuring that when the floating-point control flow diverges from the real one a warning is issued.
The proposed transformation takes into account round-off errors by abstracting the Boolean expressions in the guards of the original program. This is done by means of two Boolean abstractions $\betaPos{}, \betaNeg{}: \FBExprDom \rightarrow \FBExprDom$.

Given $\fpbool\in\FBExpr$, let $\fv{\fpbool}$ be the set of free variables in $\fpbool$. For all $\sigma: \{\real{\tilde{x}}\mid\tilde{x}\in\fv{\fpbool}\} \rightarrow \R$, $\tilde{\sigma}: \fv{\fpbool} \rightarrow \Fp$, and $\fvar\in\fv{\fpbool}$ such that $\FtoR{\tilde{\sigma}(\fvar)} = \sigma(\real{\fvar})$, $\betaPos{}$ and $\betaNeg{}$ satisfy the following properties. 
\begin{enumerate}
    \item\label{pt:pos_prop} $\evalFBExpr{\tilde{\sigma}}{\betaPos{\fpbool}} \Rightarrow
     \evalFBExpr{\tilde{\sigma}}{\fpbool} \wedge \evalBExpr{\sigma}{\FtoRB{\fpbool}}$.
    \item\label{pt:neg_prop} $\evalFBExpr{\tilde{\sigma}}{\betaNeg{\fpbool}} \Rightarrow
    \evalFBExpr{\tilde{\sigma}}{\neg\fpbool} \wedge \evalBExpr{\sigma}{\neg\FtoRB{\fpbool}}$.
\end{enumerate}
Property~\ref{pt:pos_prop} states that for all floating-point Boolean expressions $\fpbool$, $\betaPos{\fpbool}$ implies both $\fpbool$ and its real counterpart.
Symmetrically, Property~\ref{pt:neg_prop} ensures that $\betaNeg{\fpbool}$ implies both the negation of $\fpbool$ and the negation of its real counterpart.
\begin{example}
    \label{ex:beta}
The Boolean abstractions $\betaPos{}$ and $\betaNeg{}$ can be instantiated as follows for conjunctions and disjunction of sign tests.
Properties \ref{pt:pos_prop} and \ref{pt:neg_prop} are formally proven
in \PVS\ to hold for the following definitions of $\betaPos{}$ and $\betaNeg{}$.
Let $\vfpaexpr\in\FAExprDom$ and $\fperr\in \Fp$ such that $|\vfpaexpr - \FtoRA{\vfpaexpr}| \leq \fperr$.
\begin{align*}
    & \betaPos{\vfpaexpr \leq 0} = \vfpaexpr \leq -\fperr
    &&\betaNeg{\vfpaexpr \leq 0} = \vfpaexpr >      \fperr\\
    & \betaPos{\vfpaexpr \geq 0} = \vfpaexpr \geq   \fperr
    &&\betaNeg{\vfpaexpr \geq 0} = \vfpaexpr <     -\fperr\\
    & \betaPos{\vfpaexpr <    0} = \vfpaexpr <     -\fperr
    &&\betaNeg{\vfpaexpr <    0} = \vfpaexpr \geq   \fperr\\
    & \betaPos{\vfpaexpr >    0} = \vfpaexpr >      \fperr
    &&\betaNeg{\vfpaexpr >    0} = \vfpaexpr \leq  -\fperr\\
    & \betaPos{\fpbool_1 \wedge \fpbool_2} = \betaPos{\fpbool_1} \wedge \betaPos{\fpbool_2}
    &&\betaNeg{\fpbool_1 \wedge \fpbool_2} = \betaNeg{\fpbool_1} \vee \betaNeg{\fpbool_2}\\
    & \betaPos{\fpbool_1 \vee \fpbool_2} = \betaPos{\fpbool_1} \vee \betaPos{\fpbool_2}
    &&\betaNeg{\fpbool_1 \vee \fpbool_2} = \betaNeg{\fpbool_1} \wedge \betaNeg{\fpbool_2}\\
    & \betaPos{\neg\fpbool} = \betaNeg{\fpbool}
    &&\betaNeg{\neg\fpbool} = \betaPos{\fpbool}\\
\end{align*}
The abstractions performed for sign tests are not correct for generic inequalities of the form $a \leq b$.
In this case, to compensate for the round-off errors of both expressions, additional floating-point operations must be performed.
Thus, the round-off error generated by such operations needs to be considered as well to obtain a sound approximation. The naive application of this strategy leads to a non-\-terminating transformation. The design of an effective approximation for these generic inequalities is left as future work.
\end{example}

The program transformation is defined as follows.
\begin{definition}[Program Transformation]
    \label{def:prog_trans}
    Let $\fpfun(\fvar_1,\ldots,\fvar_n)=\stm \in\Prog$ be a floating-point program that does not 
    contain any $\stabWarning$ statements, the transformed program is defined as
    $\fpfun(\fvar_1,\ldots,\fvar_n)=\tauProg{\stm}$
    where $\tauProg{}$ is defined as follows.
    \begin{align*}
        &\tauProg{\fpaexpr} = \fpaexpr\\[1ex]
        &\tauProg{\ite{\fpbool}{\stm_{1}}{\stm_{2}}} =\\
        &\qquad \ite{\betaPos{\fpbool}}{\tauProg{\stm_{1}}}
                                 {\ite{\betaNeg{\fpbool}}{\tauProg{\stm_{2}}}{\stabWarning}}\\[1ex]
        &\tauProg{\tagif \fpbool_1 \tagthen \FStm_1\
        [\tagelsif \fpbool_i \tagthen \FStm_i]_{i=2}^{n-1} \tagelse \FStm_{n}} =\\
        &\qquad\tagif \betaPos{\fpbool_1} \tagthen \tauProg{\FStm_1}\
         [\tagelsif \betaPos{\fpbool_i} \wedge
         {\textstyle \bigwedge_{j = 1}^{i-1}} \betaNeg{\fpbool_j}
         \tagthen \tauProg{\FStm_i}]_{i=2}^{n-1}\\
         &\qquad\tagelsif {\textstyle \bigwedge_{j = 1}^{n-1}} \betaNeg{\fpbool_j}
         \tagthen \tauProg{\FStm_{n}}\\
         &\qquad\tagelse \stabWarning\\[1ex]
        &\tauProg{\letStm{\fvar}{\fpaexpr}{\stm}} = \letStm{\fvar}{\fpaexpr}{\tauProg{\stm}}
    \end{align*}
\end{definition}
In the case of the binary conditional statement, the \emph{then} branch of the transformed program is taken when $\betaPos{\fpbool}$ is satisfied.
By Property~\ref{pt:pos_prop}, this means that in the original program both $\fpbool$ and $\FtoR{\fpbool}$ hold and, thus, the \emph{then} branch is taken in both real and floating-point control flows.
Similarly, the \emph{else} branch of the transformed program is taken when $\betaNeg{\fpbool}$ holds.
This means, by Property~\ref{pt:neg_prop}, that in the original program the else branch is taken in both real and floating-point control flows.
In the case real and floating-flows diverge, neither $\betaPos{\fpbool}$ nor $\betaNeg{\fpbool}$ is satisfied and a warning is returned.

In the case of the n-ary conditional statements, the guard $\fpbool_i$ of the $i$-th branch is replaced by the conjunction of $\betaPos{\fpbool_i}$ and $\betaNeg{\fpbool_j}$ for all the previous branches $j<i$.
By properties~\ref{pt:pos_prop} and \ref{pt:neg_prop}, it follows that the transformed program takes the $i$-th branch only when the same branch is taken in both real and floating-point control flows of the original program.
Additionally, a warning is issued by the transformed program when real and floating-point control flows of the original program differ.

The following theorem states the correctness of the program transformation $\tauProg{}$.
If the transformed program $\tauProg{\fprog}$ returns an output
$\tilde{r}$ different from $\stabWarning$, then the original program
follows a stable path and returns the floating-point output $\tilde{r}$.
Furthermore, in the case the original program presents an unstable behavior, the transformed program returns $\stabWarning$.
\begin{theorem}[Program Transformation Correctness]
    \label{th:corr_trans}
    Given $\fpfun(\fvar_1,\!\ldots,\!\fvar_n)\!=\!\stm \in\Prog$,
    $\sigma: \{\real{\fvar_1} \dots \real{\fvar_n}\} \rightarrow \R$, and
    $\tilde{\sigma}: \{\fvar_1 \dots \fvar_n\} \rightarrow \Fp$,
    such that for all $i\in\{1,\dots,n\}$,
    $\FtoR{\tilde{\sigma}(\fvar_i)} = \sigma(\real{\fvar_i})$:
    \begin{enumerate}
    \item for all $\ct{\rcond'}{\fpcond'}{\rres'}{\fpres'}{t'}
    \in \Ssem{\tauProg{\stm}}{\botEnv}{}$
    such that $\fpres\neq\botUnstt$,
    there exists $\ct{\rcond}{\fpcond}{\rres}{\fpres}{\stt}
    \in \Ssem{\stm}{\botEnv}{}$ such that
    $\evalFBExpr{\tilde{\sigma}}{\fpcond'} \Rightarrow
    \evalBExpr{\sigma}{\rcond} \wedge \evalFBExpr{\tilde{\sigma}}{\fpcond}$
    and $\fpres = \fpres'$;
    \item for all $\ct{\rcond}{\fpcond}{\rres}{\fpres}{\unstt}
    \!\in \Ssem{\stm}{\botEnv}{}$,
    there exists $\ct{\rcond'}{\!\fpcond'}{\rres'}{\!\botUnstt}{t'}
    \!\in \Ssem{\tauProg{\stm}}{\botEnv}{}$ such that
    $\evalBExpr{\sigma}{\rcond} \wedge \evalFBExpr{\tilde{\sigma}}{\fpcond} \Rightarrow
    \evalFBExpr{\tilde{\sigma}}{\fpcond'}$.
    \end{enumerate}
\end{theorem}
The program transformation defined in \smartref{def:prog_trans} has
been formalized and  \smartref{th:corr_trans} has been proven correct
in PVS.\footnote{This formalization is available at \url{https://shemesh.larc.nasa.gov/fm/PRECiSA}.}

It is important to remark that the intended semantics of the
floating-point transformed program is the real-valued semantics of the
original one, i.e., the real-valued semantics of the transformed program is irrelevant.
Therefore, even if the transformed program presents unstable tests,
\smartref{th:corr_trans} ensures that its floating-point
control flow preserves the control flow of stable tests in the original
program.
\begin{example}
Consider the program \textit{eps\_line}, which is part of the ACCoRD conflict
detection and resolution algorithm~\cite{DowekMC05}.
This function is used to compute an implicitly coordinated horizontal resolution direction for the aircraft involved in a pair-wise conflict.
\begin{align*}
    \textit{eps\_line}(\tilde{v}_{x},\tilde{v}_{y},\tilde{s}_{x},\tilde{s}_{y})=&\tagif \widetilde{expr}> 0 \tagthen 1\ 
    \tagelsif \widetilde{expr} < 0 \tagthen -1\
    \tagelse 0,
\end{align*}
where $\widetilde{\mathit{expr}} = (\tilde{s}_{x}*\tilde{v}_{y})- (\tilde{s}_{y}*\tilde{v}_{x})$ and $\tilde{v}_{x},\tilde{v}_{y},\tilde{s}_{x},\tilde{s}_{y}$ are floating-point variables.
For example, if the values of such variables are assumed to lie in the range $[-100, 100]$, the tool \precisa{}~\cite{MoscatoTDM17,TitoloFMM18} can be used to compute the round-off error estimation $\fperr =6.4801497501321145\times{}10^{-12}$ for $\widetilde{\mathit{expr}}$.
\precisa{} is a tool that over-approximates the round-off error of floating-point programs. 
It is fully automatic and generates \PVS{} proof certificates that guarantee the correctness of the error estimations with respect to the floating-point IEEE-754 standard.
The following program is obtained by using the transformation $\tauProg{}$ with the Boolean approximations of \smartref{ex:beta}.
\begin{align*}
    \tauProg{\textit{eps\_line}(\tilde{v}_{x},\tilde{v}_{y},\tilde{s}_{x},\tilde{s}_{y})}=&\tagif \widetilde{\mathit{expr}}> \fperr \tagthen 1\ 
    \tagelsif \widetilde{\mathit{expr}} < -\fperr \tagthen -1\\
    &\quad\tagelsif \widetilde{\mathit{expr}} \geq \fperr \wedge
              \widetilde{\mathit{expr}} \leq -\fperr \tagthen 0\
    \tagelse \stabWarning
\end{align*}
The condition $\widetilde{\mathit{expr}} \geq \fperr \wedge \widetilde{\mathit{expr}} \leq -\fperr$ never holds since $\fperr$ is a positive number. Therefore, the transformed program never returns 0.
Indeed, when $\widetilde{\mathit{expr}}$ is close to $0$, the test is unstable. 
The transformed program detects these unstable cases and returns a warning.
\end{example}

\section{Case Study: PolyCARP algorithm}
\label{sec:polycarp}

PolyCARP\footnote{PolyCARP is available at \url{https://github.com/nasa/polycarp}.} (Algorithms for Computations with Polygons)~\cite{NarkawiczH16,NarkawiczMD17} is a suite of algorithms for geo-containment applications.
One of the main applications of PolyCARP is to provide geofencing capabilities to unmanned aerial systems (UAS), \ie{} detecting whether a UAS is inside or outside a given geographical region, which is modeled using a 2D polygon with a minimum and a maximum altitude.
Another application of PolyCARP is the detection of weather cells, modeled as moving polygons, along an aircraft trajectory.

A core piece of logic in PolyCARP is the polygon containment algorithm, i.e., the algorithm that checks whether or not a point lies in the interior of a polygon.
Algorithms for polygon containment have to be carefully implemented since numerical errors may lead to wrong answers, even in cases where the point is far from the boundaries of the polygon.
PolyCARP uses several techniques to detect if a point is contained in a polygon.
One of these techniques relies on the computation of the   \emph{winding number}.
This number corresponds to the number of times the polygon winds around $p$.

Consider two consecutive vertices $v$ and $v'$ of the polygon in the Cartesian plane with the point $p$ as the origin.
The function \textit{winding\_number\_edge} checks in which quadrants $v$ and $v'$ are located and counts how many axes are crossed by the edge $(v,v')$.
If $v$ and $v'$ belong to the same quadrant, the contribution of the edge to the winding number is 0 since no axis is crossed.
If $v$ and $v'$ lie in adjacent quadrants, the contribution is 1 (\resp{} -1) if moving from $v$ to $v'$ along the edge is in counterclockwise (\resp{} clockwise) direction.
In the case $v$ and $v'$ are in opposite quadrants, the determinant is computed for checking the direction of the edge. If it is counterclockwise the contribution is 2, otherwise it is -2.
The winding number is obtained as the sum of the contributions of all the edges of the polygon. If the result is 0 or 4, the point is inside the polygon, otherwise, it is outside.
\begin{align*}
    &\mathit{winding\_number\_edge}(v_x, v_y, v'_x, v'_y, p_x, p_y) =\\
    &\quad\taglet \mathit{t}_x = v_x - p_x  \tagin\ \taglet \mathit{t}_y = v_y - p_y \tagin\
    \taglet \mathit{n}_x = v'_x - p_x \tagin\ \taglet \mathit{n}_y = v'_y - p_y \tagin\\
    &\qquad\tagif \mathit{same\_quad}
       \tagthen 0 \\
    &\qquad\tagelsif\ \mathit{adj\_quad\_ctrclock}
        \tagthen 1\\
    &\qquad\tagelsif\ \mathit{adj\_quad\_clock}
        \tagthen -1 \\
    &\qquad\tagelsif\ \mathit{det\_pos}
        \tagthen 2\\
    &\qquad\tagelse\ -2
\end{align*}
where
\begin{align*}
&\mathit{same\_quad} =\\
    &\qquad\begin{aligned}[t]
    &(\mathit{t}_x \geq 0 \wedge \mathit{t}_y \geq 0 \wedge
      \mathit{n}_x \geq 0 \wedge \mathit{n}_y \geq 0) \vee
     (\mathit{t}_x \leq 0 \wedge \mathit{t}_y \geq 0 \wedge
      \mathit{n}_x \leq 0 \wedge \mathit{n}_y \geq 0)\ \vee\\
    &(\mathit{t}_x \geq 0 \wedge \mathit{t}_y \leq 0 \wedge
      \mathit{n}_x \geq 0 \wedge \mathit{n}_y \leq 0) \vee
     (\mathit{t}_x \leq 0 \wedge \mathit{t}_y \leq 0 \wedge
      \mathit{n}_x \leq 0 \wedge \mathit{n}_y \leq 0)
    \end{aligned}\\
&\mathit{adj\_quad\_ctrclock} =\\
    &\qquad\begin{aligned}[t]
       &(\mathit{t}_x \geq 0 \wedge \mathit{t}_y \leq 0 \wedge
         \mathit{n}_x \geq 0 \wedge \mathit{n}_y \geq 0)\vee
        (\mathit{t}_x \geq 0 \wedge \mathit{t}_y \geq 0 \wedge
         \mathit{n}_x \leq 0 \wedge \mathit{n}_y \geq 0)\ \vee\\
       &(\mathit{t}_x \leq 0 \wedge \mathit{t}_y \geq 0 \wedge
         \mathit{n}_x \leq 0 \wedge \mathit{n}_y \leq 0) \vee
        (\mathit{t}_x \leq 0 \wedge \mathit{t}_y \leq 0 \wedge
         \mathit{n}_x \geq 0 \wedge \mathit{n}_y \leq 0),
    \end{aligned}\\
&\mathit{adj\_quad\_clock} =\\
    &\qquad\begin{aligned}[t]
        &(\mathit{t}_x \geq 0 \wedge \mathit{t}_y \geq 0 \wedge
          \mathit{n}_x \geq 0 \wedge \mathit{n}_y \leq 0) \vee
         (\mathit{t}_x \leq 0 \wedge \mathit{t}_y \geq 0 \wedge
          \mathit{n}_x \leq 0 \wedge \mathit{n}_y \geq 0)\ \vee\\
        &(\mathit{t}_x \leq 0 \wedge \mathit{t}_y \leq 0 \wedge
          \mathit{n}_x \leq 0 \wedge \mathit{n}_y \geq 0) \vee
         (\mathit{t}_x \geq 0 \wedge \mathit{t}_y \leq 0 \wedge
          \mathit{n}_x \leq 0 \wedge \mathit{n}_y \leq 0),
    \end{aligned}\\
&\mathit{det\_pos} = (\mathit{n}_x - \mathit{t}_x) * \mathit{t}_y
              - (\mathit{n}_y - \mathit{t}_y) * \mathit{t}_x \leq 0.
\end{align*}
The function \textit{winding\_number\_edge} has been verified in \PVS{} using real arithmetic.
However, due to floating-point errors, taking the incorrect branch for one of the edges in the computation of the winding number may result in an incorrect conclusion about the position of the point \wrt{} the polygon.
In order to overcome this problem, the transformation $\tauProg{}$ of \smartref{def:prog_trans} is applied to the function $\mathit{winding\_number\_edge}$ resulting in the following function.
Given initial bounds for the input variables, \precisa{}~\cite{MoscatoTDM17,TitoloFMM18} can be used to compute the round-off error estimations for $n_x$, $n_y$, $t_x$, $t_y$ and the determinant, which are denoted $\fperr_{t_x}$, $\fperr_{t_y}$, $\fperr_{n_x}$, $\fperr_{n_y}$, and $\fperr_{\textit{det}}$, respectively.
\begin{align*}
    &\tauProg{\mathit{winding\_number\_edge}(v_x, v_y, v'_x, v'_y, p_x, p_y)} =\\
    &\quad\taglet \mathit{t}_x = v_x - p_x  \tagin\ \taglet \mathit{t}_y = v_y - p_y \tagin\
    \taglet \mathit{n}_x = v'_x - p_x \tagin\ \taglet \mathit{n}_y = v'_y - p_y \tagin\\
    &\qquad\tagif\ \mathit{same\_quad^{\beta}}
       \tagthen 0 \\
    &\qquad\tagelsif\ \mathit{adj\_quad\_ctrclock^{\beta}}
        \tagthen 1\\
    &\qquad\tagelsif\ \mathit{adj\_quad\_clock^{\beta}}
        \tagthen -1 \\
    &\qquad\tagelsif\ \mathit{det\_pos^{\beta}}
        \tagthen 2\\
    &\qquad\tagelsif\ \mathit{original\_else^{\beta}} \tagelse -2\\
    &\qquad\tagelse\ \stabWarning,
\end{align*}
where
\begin{align*}
&\mathit{same\_quad^{\beta}} = \betaPos{\mathit{same\_quad}} =
   \begin{aligned}[t]
    &(\mathit{t}_x \geq  \fperr_{t_x} \wedge \mathit{t}_y \geq  \fperr_{t_y} \wedge
      \mathit{n}_x \geq  \fperr_{n_x} \wedge \mathit{n}_y \geq  \fperr_{n_y})\ \vee\\
    &(\mathit{t}_x \leq -\fperr_{t_x} \wedge \mathit{t}_y \geq  \fperr_{t_y} \wedge
      \mathit{n}_x \leq -\fperr_{n_x} \wedge \mathit{n}_y \geq  \fperr_{n_y})\ \vee\\
    &(\mathit{t}_x \geq  \fperr_{t_x} \wedge \mathit{t}_y \leq -\fperr_{t_y} \wedge
      \mathit{n}_x \geq  \fperr_{n_x} \wedge \mathit{n}_y \leq -\fperr_{n_y})\ \vee\\
    &(\mathit{t}_x \leq -\fperr_{t_x} \wedge \mathit{t}_y \leq -\fperr_{t_y} \wedge
      \mathit{n}_x \leq -\fperr_{n_x} \wedge \mathit{n}_y \leq -\fperr_{n_y}),
    \end{aligned}\\[1ex]
&\mathit{adj\_quad\_ctrclock^{\beta}} =
    \betaPos{\mathit{adj\_quad\_counterclock}}
    \wedge
    \betaNeg{\mathit{same\_quad}},\\[1ex]
&\mathit{adj\_quad\_clock^{\beta}} =
    \begin{aligned}[t]
        &\betaPos{\mathit{adj\_quad\_clock}}
        \wedge
        \betaNeg{\mathit{adj\_quad\_ctrclock}}
        \ \wedge\\
        & \betaNeg{\mathit{same\_quad}},\\[1ex]
    \end{aligned}\\
&\mathit{det\_pos^{\beta}} =
    \begin{aligned}[t]
        & (\mathit{n}_x - \mathit{t}_x) * \mathit{t}_y
        - (\mathit{n}_y - \mathit{t}_y) * \mathit{t}_x \leq -\fperr_{\mathit{det}}
        \wedge
        \betaNeg{\mathit{adj\_quad\_clock}}
        \ \wedge\\
        &
        \betaNeg{\mathit{adj\_quad\_ctrclock}}
        \wedge
        \betaNeg{\mathit{same\_quad}},\\[1ex]
    \end{aligned}\\
&\mathit{original\_else^{\beta}} =
    \begin{aligned}[t]
        & (\mathit{n}_x - \mathit{t}_x) * \mathit{t}_y
        - (\mathit{n}_y - \mathit{t}_y) * \mathit{t}_x > \fperr_{\mathit{det}}
        \wedge
        \betaNeg{\mathit{adj\_quad\_clock}}
        \ \wedge\\
        &
        \betaNeg{\mathit{adj\_quad\_ctrclock}}
        \wedge
        \betaNeg{\mathit{same\_quad}},\\[1ex]
    \end{aligned}\\[1ex]
&\betaNeg{\mathit{same\_quad}} =
    \begin{aligned}[t]
    &(\mathit{t}_x < -\fperr_{t_x} \vee \mathit{t}_y < -\fperr_{t_y} \vee
      \mathit{n}_x < -\fperr_{n_x} \vee \mathit{n}_y < -\fperr_{n_y})\ \wedge\\
    &(\mathit{t}_x >  \fperr_{t_x} \vee \mathit{t}_y < -\fperr_{t_y} \vee
      \mathit{n}_x >  \fperr_{n_x} \vee \mathit{n}_y < -\fperr_{n_y})\ \wedge\\
    &(\mathit{t}_x < -\fperr_{t_x} \vee \mathit{t}_y >  \fperr_{t_y} \vee
      \mathit{n}_x < -\fperr_{n_x} \vee \mathit{n}_y >  \fperr_{n_y})\ \wedge\\
    &(\mathit{t}_x >  \fperr_{t_x} \vee \mathit{t}_y >  \fperr_{t_y} \vee
      \mathit{n}_x >  \fperr_{n_x} \vee \mathit{n}_y >  \fperr_{n_y}),
    \end{aligned}\\[1ex] 
&\betaPos{\mathit{adj\_quad\_ctrclock}} =
    \begin{aligned}[t]
       &(\mathit{t}_x \geq  \fperr_{t_x} \wedge \mathit{t}_y \leq -\fperr_{t_y} \wedge
         \mathit{n}_x \geq  \fperr_{n_x} \wedge \mathit{n}_y \geq
         \fperr_{n_y})\ \vee\\
       &(\mathit{t}_x \geq  \fperr_{t_x} \wedge \mathit{t}_y \geq  \fperr_{t_y} \wedge
         \mathit{n}_x \leq -\fperr_{n_x} \wedge \mathit{n}_y \geq
         \fperr_{n_y})\ \vee\\
       &(\mathit{t}_x \leq -\fperr_{t_x} \wedge \mathit{t}_y \geq  \fperr_{t_y} \wedge
         \mathit{n}_x \leq -\fperr_{n_x} \wedge \mathit{n}_y \leq -\fperr_{n_y})\ \vee\\
       &(\mathit{t}_x \leq -\fperr_{t_x} \wedge \mathit{t}_y \leq -\fperr_{t_y} \wedge
         \mathit{n}_x \geq  \fperr_{n_x} \wedge \mathit{n}_y \leq -\fperr_{n_y}),
    \end{aligned}\\[1ex]
&\betaNeg{\mathit{adj\_quad\_ctrclock}}=
    \begin{aligned}[t]
       &(\mathit{t}_x < -\fperr_{t_x} \vee \mathit{t}_y >  \fperr_{t_y} \vee
         \mathit{n}_x < -\fperr_{n_x} \vee \mathit{n}_y <
         -\fperr_{n_y})\ \wedge\\
       &(\mathit{t}_x < -\fperr_{t_x} \vee \mathit{t}_y < -\fperr_{t_y} \vee
         \mathit{n}_x >  \fperr_{n_x} \vee \mathit{n}_y <
         -\fperr_{n_y})\ \wedge\\
       &(\mathit{t}_x >  \fperr_{t_x} \vee \mathit{t}_y < -\fperr_{t_y} \vee
         \mathit{n}_x >  \fperr_{n_x} \vee \mathit{n}_y >
         \fperr_{n_y})\ \wedge\\
       &(\mathit{t}_x >  \fperr_{t_x} \vee \mathit{t}_y >  \fperr_{t_y} \vee
         \mathit{n}_x < -\fperr_{n_x} \vee \mathit{n}_y >  \fperr_{n_y}),
    \end{aligned}\\[1ex]
&\betaPos{\mathit{adj\_quad\_clock}} =
    \begin{aligned}[t]
        &(\mathit{t}_x \geq  \fperr_{t_x} \wedge \mathit{t}_y \geq  \fperr_{t_y} \wedge
          \mathit{n}_x \geq  \fperr_{n_x} \wedge \mathit{n}_y \leq
          -\fperr_{n_y})\ \vee\\
        &(\mathit{t}_x \leq -\fperr_{t_x} \wedge \mathit{t}_y \geq  \fperr_{t_y} \wedge
          \mathit{n}_x \leq -\fperr_{n_x} \wedge \mathit{n}_y \geq
          \fperr_{n_y})\ \vee\\
        &(\mathit{t}_x \leq -\fperr_{t_x} \wedge \mathit{t}_y \leq -\fperr_{t_y} \wedge
          \mathit{n}_x \leq -\fperr_{n_x} \wedge \mathit{n}_y \geq  \fperr_{n_y})\ \vee\\
        &(\mathit{t}_x \geq  \fperr_{t_x} \wedge \mathit{t}_y \leq -\fperr_{t_y} \wedge
          \mathit{n}_x \leq -\fperr_{n_x} \wedge \mathit{n}_y \leq -\fperr_{n_y}),
    \end{aligned}\\[1ex]
&\betaNeg{\mathit{adj\_quad\_clock}} =
    \begin{aligned}[t]
        &(\mathit{t}_x < -\fperr_{t_x} \vee \mathit{t}_y < -\fperr_{t_y} \vee
          \mathit{n}_x < -\fperr_{n_x} \vee \mathit{n}_y >
          \fperr_{n_y})\ \wedge\\
        &(\mathit{t}_x >  \fperr_{t_x} \vee \mathit{t}_y < -\fperr_{t_y} \vee
          \mathit{n}_x >  \fperr_{n_x} \vee \mathit{n}_y <
          -\fperr_{n_y})\ \wedge\\
        &(\mathit{t}_x >  \fperr_{t_x} \vee \mathit{t}_y >  \fperr_{t_y} \vee
          \mathit{n}_x >  \fperr_{n_x} \vee \mathit{n}_y < -\fperr_{n_y})\ \wedge\\
        &(\mathit{t}_x < -\fperr_{t_x} \vee \mathit{t}_y >  \fperr_{t_y} \vee
          \mathit{n}_x >  \fperr_{n_x} \vee \mathit{n}_y >  \fperr_{n_y}).
    \end{aligned}
\end{align*}
Consider a polygonal geofence and a set of randomly generated points in the square that circumscribes it.
For each edge of the polygon and each generated point, the original function $\mathit{winding\_number\_edge}$ is executed by using both exact real arithmetic and double-precision floating-point arithmetic. Additionally, the transformed function $\tauProg{\mathit{winding\_number\_edge}}$ is executed with double-precision floating-point arithmetic.
For these randomly generated points, both the original and the transformed program return the same result.
However, the closer the generated point is to the border of the polygon, the more likely is for the original program to take an unstable path.
By considering a set of randomly generated points very close to the edges of the polygon, the transformed program always returns a warning, showing that these are the cases for which the floating-point computation may diverge from the real one.
Since an over-approximation of the round-off error is used, not all the generated warnings reflect an actual problem.
In fact, false warnings occur when the compensated error computed by the abstraction is larger than the round-off error that actually occurs in the computation.
The amount of false warnings converges to the $50\%$ of the number of total warnings as the distance to the edge decreases.

\section{Related Work}
\label{sec:related}

Recently, several program transformations have been proposed with the aim of improving accuracy and efficiency of floating-point computations. It is possible to distinguish two kinds of approaches: precision allocation tools and program optimization ones.
Precision allocation (or tuning) tools aim at selecting the lowest floating-point precision that is necessary to achieve a desired accuracy. This approach avoids using more precision than needed and improves the performance of the program.
Rosa~\cite{DarulovaK14,DarulovaK17} uses a compilation algorithm that, from an ideal real-valued implementation, produces a finite-precision version (if it exists) that is guaranteed to meet the desired overall precision. Rosa soundly deals with unstable tests and with bounded loops.
Similarly, \FPTuner~\cite{ChiangBBSGR17} implements a rigorous approach to precision allocation of mixed-precision arithmetic expressions.
Precimonius~\cite{Rubio-GonzalezNNDKSBIH13} is a dynamic tool able to identify parts of a program that can be performed at a lower precision. It generates a transformed program where each floating-point variable is typed to the lowest precision necessary to meet a set of given accuracy and performance constraints. Hence, the transformed program uses variables of lower precision and performs better than the original program.

Program optimization tools aim at improving the accuracy of floating-point programs by rewriting arithmetic expressions in equivalent ones with a lower accumulated round-off error.
Herbie~\cite{PanchekhaSWT15} is a tool that automatically improves the accuracy of floating-point programs though a heuristic search. Herbie detects the expressions where rounding-errors occur and it applies a series of rewriting and simplification rules. It generates a set of transformed programs that are equivalent to the original one but potentially more accurate. The rewriting and simplification process is then applied recursively to the generated transformed programs until the most accurate program is obtained.
CoHD~\cite{ThevenouxLM15} is a source-to-source transformer for C code that automatically compensates for the round-off errors of some basic floating-point operations.
SyHD~\cite{ThevenouxLM17} is a C code optimizer that explores a set of programs generated by CoDH and selects the one with the best accuracy and computation-time trade-off.
The tool Sardana~\cite{IoualalenM13}, given a Lustre~\cite{CaspiPHP87} program, produces a set of equivalent programs with simplified arithmetic expressions. Then, it selects the ones for which a better accuracy bound can be proved.
Salsa~\cite{DamoucheMC17c} combines Sardana with techniques for intra-procedure~\cite{DamoucheMC15} and inter-procedure~\cite{DamoucheMC17a,DamoucheMC17b} program transformation in order to improve the accuracy of a target variable in larger pieces of code containing assignments and control structures.
To the best of the authors' knowledge, the program transformation proposed in this work is the only approach that addresses the problem of conditional instability for floating-point programs.

\section{Conclusion}
\label{sec:concl}

This paper presents a formally verified program transformation to detect instability in floating-point programs.
The transformed program is guaranteed to return a warning when real and floating-point flows may diverge. Otherwise, it behaves as the original program when real and floating-point control flows coincide.
The proposed approach is parametric \wrt{} two Boolean expression abstractions that return more restrictive Boolean conditions using an over-approximation of the round-off error occurring in the guard.
These abstractions cause a loss of precision since the guards occurring in the transformed program are more restrictive and, therefore, some stable original traces may be lost in the transformed program.
This leads to the possibility of having false instability warnings.
However, it is ensured that all the unstable paths of the original program are detected.

This transformation has been formalized and formally proven correct in the interactive theorem prover \PVS.
The PVS tool PVSio can be used to execute the program
transformation. However, a full integration with \precisa{} is the missing step to compute the round-off error approximations and to make the presented approach fully automatic.

The program transformation presented in this paper is the first step towards the much broader goal of improving the quality and reliability of floating-point programs.
Future work includes the extension of the formalization to a more expressive language where conditionals are allowed inside Boolean expressions and function calls and loops are supported.
This extension is not straightforward since it involves several
changes in the formalization. In fact, in such setting, the evaluation
of the expressions in the guards can also present unstable behaviors. 
Additionally, an extensive experimental evaluation is needed in order to assess the quality of the approach and its applicability to real-world applications.
Another interesting future direction is the integration of the proposed approach with tools such as Salsa~\cite{DamoucheMC17c} and Herbie~\cite{PanchekhaSWT15}.
This integration will improve the accuracy of the mathematical expressions used inside a program and, at the same time, prevent unstable tests that may cause unexpected behaviors.

\bibliographystyle{splncs}
\bibliography{biblio}

\end{document}